\documentclass{mn2e}
\newcommand{\kms}{\,\mathrm{km}\,{\mathrm{s}^{-1}}}
\newcommand{\ha  }{\hbox{$\hbox{H\,$\alpha$}$}}
\newcommand{\hb  }{\hbox{$\hbox{H\,$\beta $}$}}
\newcommand{\hei }{\hbox{$\hbox{He\,{\sc i} }$}}
\newcommand{\heii}{\hbox{$\hbox{He\,{\sc ii}}$}}
\newcommand{\feii}{\hbox{$\hbox{Fe\,{\sc ii}}$}}

\title{Optical spectroscopy of the dwarf nova U Geminorum}
\author[Unda-Sanzana$^1$,  E.,
        Marsh$^2$,         T. R.,
        Morales-Rueda$^3$, L.]{
        E.\ Unda-Sanzana, T.\,R.\ Marsh, L.\ Morales-Rueda \\
        $^1$ Instituto de Astronom\'ia,
        Universidad Cat\'olica del Norte, Antofagasta, Chile; eundas@almagesto.org\\
        $^2$ Dept. of Physics, University of Warwick,
        Coventry, United Kingdom CV4 7AL; t.r.marsh@warwick.ac.uk\\
        $^3$ Dept. of Astrophysics, University of Nijmegen,
        6500 GL Nijmegen, The Netherlands; lmr@astro.ru.nl\\
}
\date{Accepted .
      Received ;
      in original form }
\pagerange{\pageref{firstpage}--\pageref{lastpage}}
\pubyear{2006}
\usepackage{graphics}
\usepackage{graphicx}
\begin{document}
\maketitle
\label{firstpage}
\begin{abstract}
  The dwarf nova U~Gem is unique in having a direct measurement of the
  $K$-velocity of its white dwarf from \textit{HST} spectra ($K_1 = 107 \pm 2
  \kms$, Long et al. 1999). We present high-resolution optical spectra of the
  dwarf nova U~Gem in quiescence taken to test the accuracy to which the
  \textit{HST} value can be recovered from optical data. We find that, even with
  data of very high signal-to-noise ratio on this archetypal system, we cannot
  recover Long et al.'s value to better than about $20$\% by any method.
  Contamination by neighbouring emission lines seems a likely culprit.  Our data
  reveal a number of new features: Doppler tomograms show emission at low
  velocity, close to the centre of mass, and a transient, narrow absorption
  feature is seen in the Balmer lines near the line centres at the time of
  eclipse.  We suggest that stellar prominences, as previously invoked for the
  dwarf novae IP~Peg and SS~Cyg in outburst, may explain both of these features.
  The \heii 4686.75 \AA\ line emission is dominated by the gas stream/disc
  impact region. Two distinct spots are seen in Doppler maps, the first being
  very narrow and showing a velocity close to that of the accretion disc in the
  impact region, and the second much broader and located between the velocities
  of the (ballistic) stream and the (Keplerian) disc.  We present tentative
  evidence of weak spiral structure, which may support explanations for ``spiral
  shocks'' based upon 3-body effects. We find no evidence of stream-disc
  overflow in the system.
  Our data suggests an inclination angle $> 70^{\circ}$, adding to the evidence
  supporting the existence of a puzzle in the mass of U~Gem's white dwarf.
  The mass donor is clearly seen in the Doppler maps,
  with emission concentrated towards its poles, and mainly on the side facing
  the white dwarf. This suggests irradiation with shielding by the disc from
  which we estimate an $H/R$ ratio between 0.15 and 0.25.
\end{abstract}
\section{Introduction}
Theoretical models of the evolution of cataclysmic variables (CVs)
make predictions about the distributions of their physical parameters
such as orbital period, mass ratio and the individual masses of the
two stars. If these parameters can be measured, the (many) assumptions
that go into the theory can be tested. These assumptions are of
interest in the broader context of binary star evolution. The same
parameters are needed to gain an accurate understanding of the
accretion processes which dominate these stars.  Unfortunately, the
only reliable physical parameter that can be measured for the majority
of CVs is their orbital period, with perhaps the spectral type of the
donor star coming a distant second. In this ranking, the masses of the
white dwarf and the donor star are so uncertain that they are often
barely mentioned in the context of evolution. This is partly because
of the difficulty of measuring orbital inclinations common to all
forms of close binary, but also because it is hard to measure the 
the radial velocities of the binary components.

The optical spectra of most CVs are dominated by emission from
accreting material, rather than the white dwarf.  In the case of
non-magnetic systems, the emission lines are formed in discs, and
often have a double-peaked shape resulting from the Doppler shift of
matter rotating in a Keplerian disc. The high-velocity wings of the
emission lines are expected to form close to the white dwarf, where it
is hoped the flow is axi-symmetric. If so, these wings should trace
the motion of the white dwarf. This is the idea behind many studies of
CVs. However, this method fails in practice: in those systems in which
one has independent knowledge of the orbital phase, the emission line
radial velocities invariably fail to match expectations (Stover
1981). Nevertheless, the white dwarf velocities measured in this way
are still used for mass determination, perhaps because there has not
been any direct evidence for the magnitude of the distortion. This
changed after Long \& Gilliland (1999) measured directly the radial
velocity semi-amplitude of the white dwarf in U~Gem from narrow
photospheric lines visible in the ultraviolet. In this case, the phase
agreed perfectly with the predicted conjunction phase.  
Long \& Gilliland (1999) measured $K_1 = 107 \pm 2 \kms$. The direct nature
of this measurement makes it the most accurate to date for any CV, and
gives us the chance to test what the nature of the distortion of the
emission line measurements in one system at least. One other CV for
which we have a good estimate of $K_1$ is AE~Aqr (Welsh et al
1995). In this case $K_1$ is measured from the pulsations of the white
dwarf. However, AE~Aqr is a very unusual CV with flaring emission
lines that are not suited to testing the general method of measuring
$K_1$ from line emission.

U~Gem is interesting from several other points of view: it is a
double-lined, partially-eclipsing system, and is one of the brightest
dwarf novae. It has prominent emission from the gas stream/disc impact
region which has a velocity in-between that of a Keplerian disc and a
ballistic stream (Marsh et al. 1990). Finally, during outburst it has
shown spiral shocks (Groot 2001). We will see that our new data has
something to tell us on all of these issues.

\section{Observations and reduction}
The observations were taken in January 2001 at La Palma (see Table
\ref{tab:data} for details). The 4.2-m William Herschel Telescope
(WHT) was used with the double-beamed ISIS spectrograph. Two datasets,
covering wavelength ranges around H$\alpha$ and H$\beta$ were acquired
using the highest resolution gratings. The spectrograph's slit was
oriented at a position angle of $150.6^{\circ}$ to include a
comparison star for slit loss corrections.

The spectrophotometric standard HD19445 and one featureless star were
observed during the nights with best seeing ($\sim$0.7") for flux
calibration and to remove the effect of telluric lines on the red data
(Bessell 1999). Flat-fields and comparison arc spectra were taken
approximately every 60 minutes for the red and the blue dataset.  The
weather was clear throughout most of our run with seeing of order 1
arcsecond for most of it.

\begin{table*}
\caption{Summary of the observations. In this table MD stands for
'mean dispersion'. 
%
T is the mean exposure time per frame and DT is the average dead time between exposures. N is the
number of spectra obtained per night per arm.}
\begin{tabular}{llccccccrrr}
\multicolumn{1}{l}{CCD}                &
\multicolumn{1}{l}{Grating}            &
\multicolumn{1}{c}{Date}               &
\multicolumn{1}{c}{Start - End}        &
\multicolumn{1}{c}{Orbits}             &
\multicolumn{1}{c}{$\lambda$ range}   &
\multicolumn{1}{c}{MD}                 &
\multicolumn{1}{c}{FWHM}               &
\multicolumn{1}{c}{T/DT}               &
\multicolumn{1}{c}{N}                  \\
\multicolumn{1}{l}{}                   &
\multicolumn{1}{l}{}                   &
\multicolumn{1}{c}{}                   &
\multicolumn{1}{c}{(UT)}               &
\multicolumn{1}{c}{covered}            &
\multicolumn{1}{c}{(\AA)}              &
\multicolumn{1}{c}{(\AA\,pixel$^{-1}$)}&
\multicolumn{1}{c}{(\AA)}              &
\multicolumn{1}{c}{(s)}                &
\multicolumn{1}{c}{}                   \\
& & & & & & & &\\
EEV12&H2400B&12/13 Jan 2001&22:14-05:32&1.719&4618-4985&0.21&0.42&120/12&182\\
TEK4 &R1200R&12/13 Jan 2001&22:14-05:32&1.719&6343-6751&0.40&0.80&  50/6&419\\
EEV12&H2400B&13/14 Jan 2001&21:08-03:54&1.594&4618-4985&0.21&0.42&120/12&176\\
TEK4 &R1200R&13/14 Jan 2001&21:08-03:54&1.594&6343-6751&0.40&0.80&  50/6&410\\
EEV12&H2400B&14/15 Jan 2001&21:44-02:07&1.032&4618-4985&0.21&0.42&120/12&112\\
TEK4 &R1200R&14/15 Jan 2001&21:44-02:07&1.032&6343-6751&0.40&0.80&  50/6&259\\
EEV12&H2400B&15/16 Jan 2001&21:22-04:49&1.755&4618-4985&0.21&0.42&120/12&191\\
TEK4 &R1200R&15/16 Jan 2001&21:22-04:49&1.755&6343-6751&0.40&0.80&  50/6&442\\

\end{tabular}
\label{tab:data}
\end{table*}

The spectra were optimally extracted (Marsh 1989), with flat fields
interpolated from the many taken during the night. The wavelength
scales were interpolated from the nearest two arc spectra. Slit losses
were corrected using the ratio of the spectra of the comparison star
to a spectrum taken with a wide slit close to the zenith.

\section{Results}

\subsection{Average spectrum}

\begin{figure*}
\begin{center}
\hspace*{\fill}
\includegraphics[width=2.5in,angle=270]{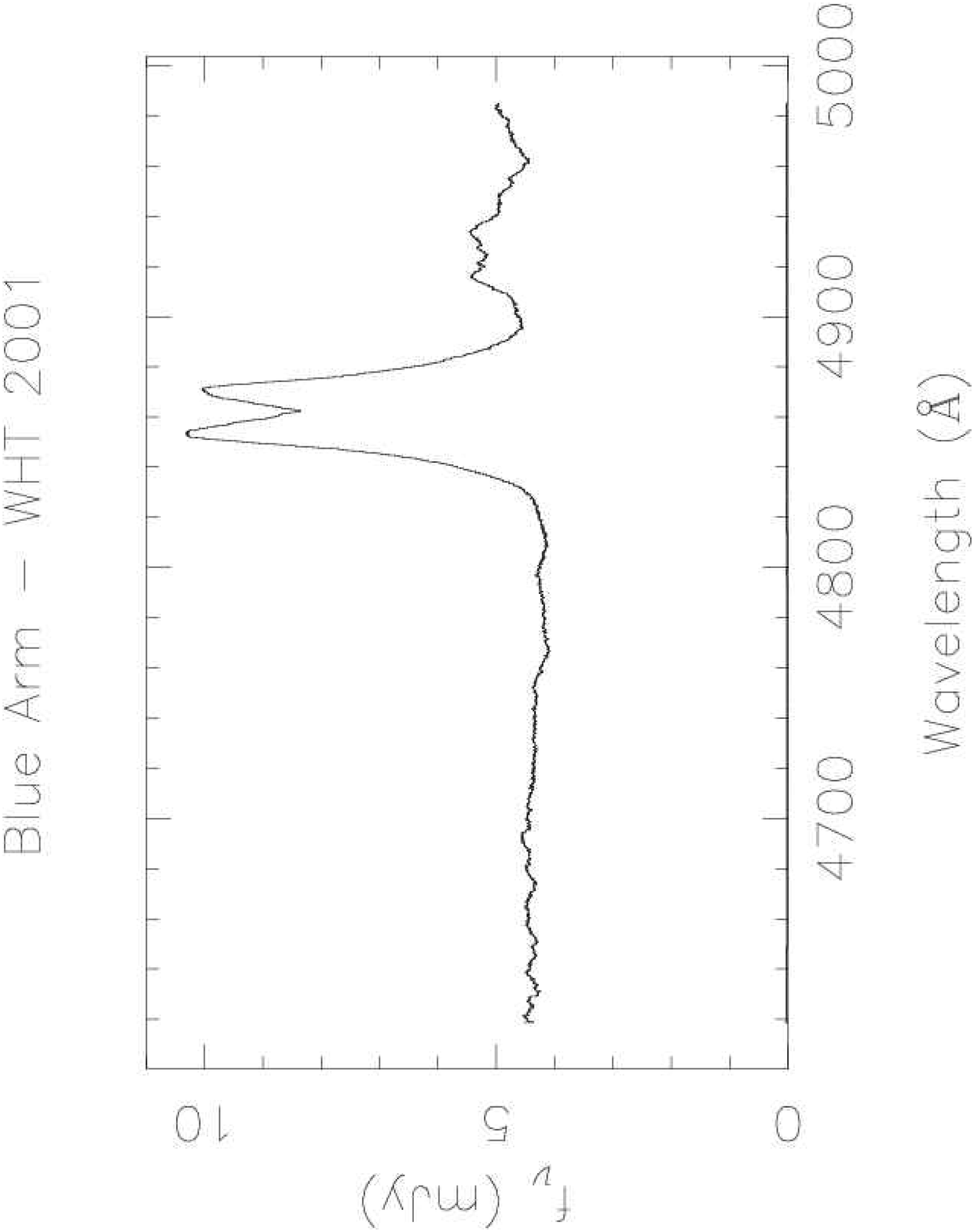}
\hspace*{\fill}
\includegraphics[width=2.5in,angle=270]{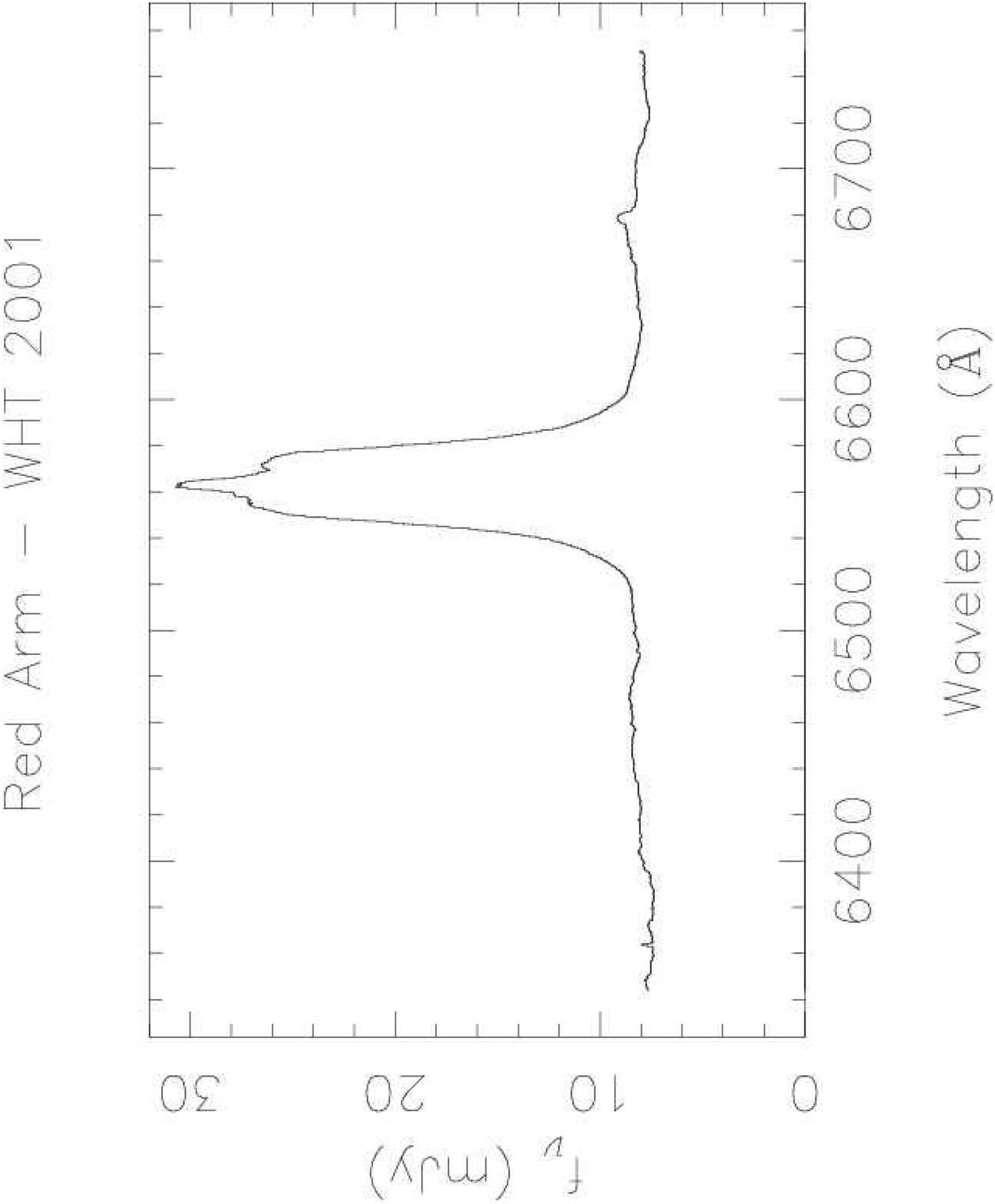}
\hspace*{\fill}
\caption{\small The average spectrum of U~Gem. Note the change in the
  vertical scale between the two wavelength ranges.}
\label{fig:av_spectrum}
\end{center}
\end{figure*}

Fig.~\ref{fig:av_spectrum} shows the average spectrum of U~Gem during
our run. U~Gem shows broad Balmer (\ha, \hb) and \feii\ 4923.92 \AA\
(hereafter \feii) emission. \hei\ 6678.149 \AA\ (hereafter \hei) and
the high excitation line \heii\ 4685.750 \AA\ (hereafter \heii) are
also present, the latter only obvious in trailed spectra however. Next
to \feii\ 4923.92 there is emission at an unidentified wavelength
(\feii\ 5018.44 \AA or \hei\ 5015), which, as we will see later, introduces some
contamination in \feii.

The $B$-band line profiles show the double-peaked structure often seen in
CVs, and attributed to the presence of a rotating accretion disc
around the compact object. The $R$-band profiles exhibit a central
stationary component which is reminiscent of similar features seen in
a handful of other systems. We discuss this component in detail in
Section \ref{sec:low_velocity_emission}.
%
%
%

\subsection{Light curves} \label{sec:light_curves}

\begin{figure*}
\begin{center}
\includegraphics[width=5.5in,angle=270]{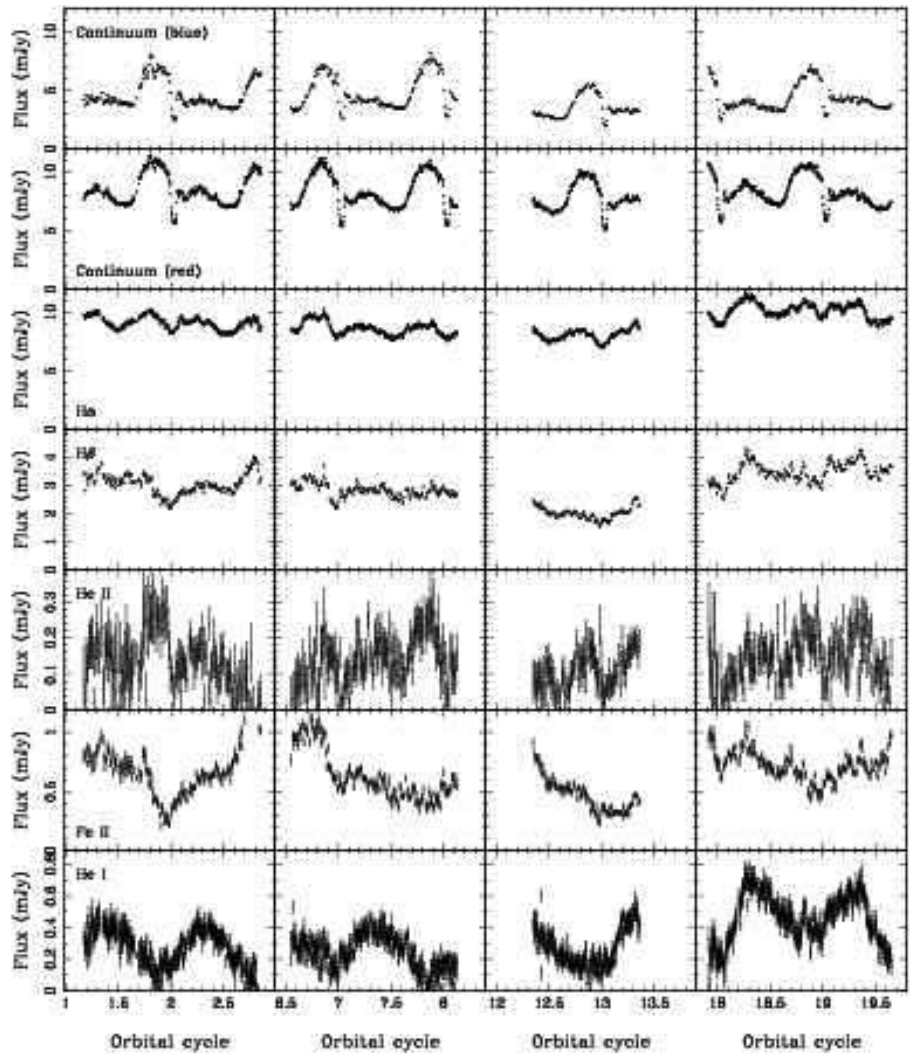}
\caption{\small Continuum and emission light curves of the WHT/ISIS
data.  The zero-point of the phase is cycle 80740 on our ephemeris.}
\label{fig:unbinned_light_curves}
\end{center}
\end{figure*}

We computed light curves of sections of the continuum, avoiding
emission lines (6400-6480 \AA, 4620-4650 \AA\ and 4720-4470 \AA),
and also of emission lines after continuum subtraction, summing
the flux within  $2000\,\kms$ of the line centre. Exceptions were made
for \feii, for which we used the range from $-1500$ to $+1500\,\kms$ in
order to prevent contamination from the wings of \hb, and \heii\ for
which we used $\pm 650 \,\kms$ due to the narrowness of the
emission.  The results are plotted in Figure~\ref{fig:unbinned_light_curves}.
We used the ephemeris of Marsh et al. (1990) updated from a
recent measurement of the inferior conjunction of the red star (Naylor et al. 2005) so that zero phase occurs at TDB
$=2451915.8618\pm0.0002$.

Both the blue and the red continuum light curves display smooth,
moderately deep, narrow eclipses just after phase 0. This is the
eclipse of the gas stream/disc impact region, or ``bright-spot'' at
the edge of the disc. The white dwarf is not eclipsed in U~Gem

The light curves of \ha\ and \hb\ show broad humps around phases 0.25
and 0.75. The humps are more marked in \ha, which makes us think that
they are largely due to ellipsoidal modulations produced by
the Roche deformation of the mass donor, which we expect to be stronger
in red light. The more complex structure of the hump in 0.75 is expected
if that one is caused by a mixture of ellipsoidal modulations and variations
in the strength of emission from the bright-spot.
%
The light curves
also display a shallow eclipse roughly centred around phase 0, and
extending symmetrically $\sim$0.1 in phase to both sides. This is
produced by the mass donor passing in front of the outer part of the
accretion disc.
%

In \heii\ the eclipse occurs slightly after phase 0. We will see in
the following sections that this is because \heii\ is dominated by
emission from the bright-spot. The peak of light near phase 0.9 is
consistent with the orbital humps detected in the continua. By
contrast, the Balmer lines are hardly eclipsed indicative of a
relatively small contribution from the bright-spot to these lines.

The light curve of \hei is strongly modulated, with the peak flux at
phase 0.3-0.4 caused by an increase in the contribution of the
bright-spot (discussed in the following sections).

\subsection{Trailed spectra}
\label{sec:trailed_spectra}

\begin{figure*}
\begin{center}
\includegraphics[width=2.4in,angle=270]{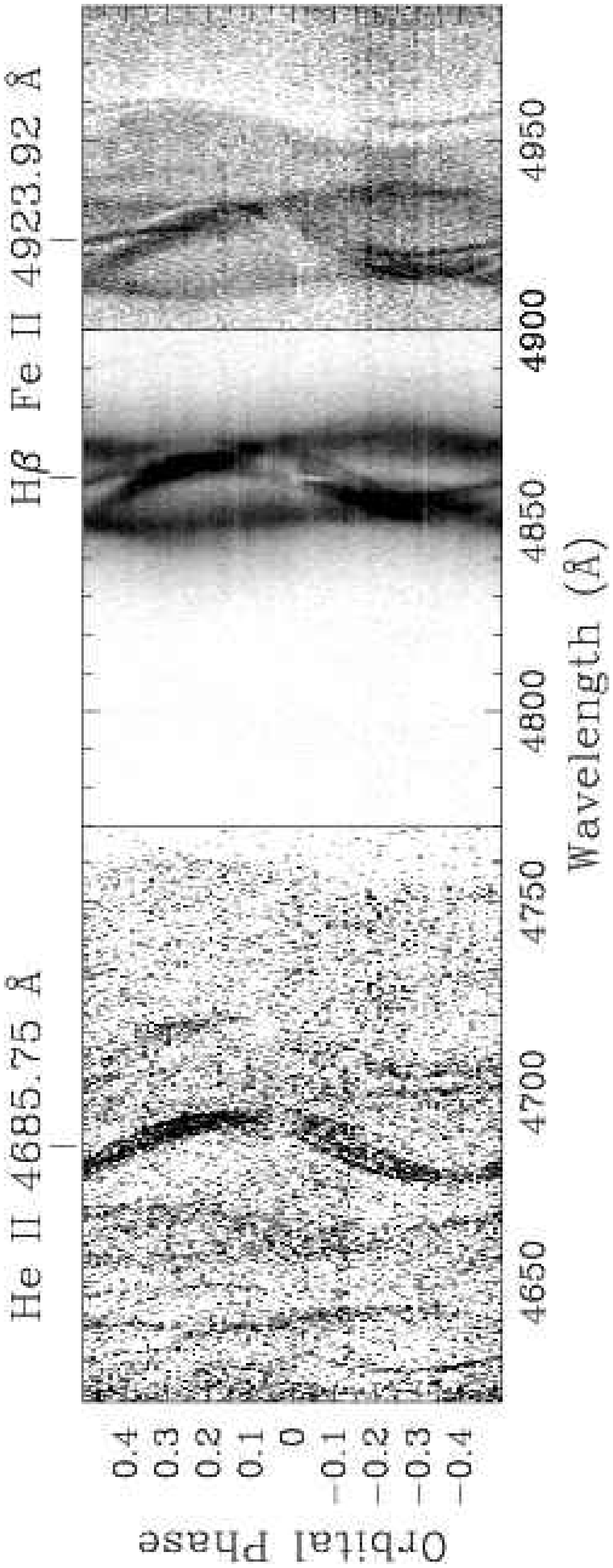}
\caption{\small The blue spectra binned into 100 bins of orbital
phase. To cope with the large dynamic range of the features 
we use 3 different normalisations running from the continuum
level to 0.5, 7.0 and 2.0 mJy above continuum for \feii,
\heii\ and \feii\ respectively.}
\label{fig:bluedata2001}
\end{center}
\end{figure*}

\begin{figure*}
\begin{center}
\includegraphics[width=2.4in,angle=270]{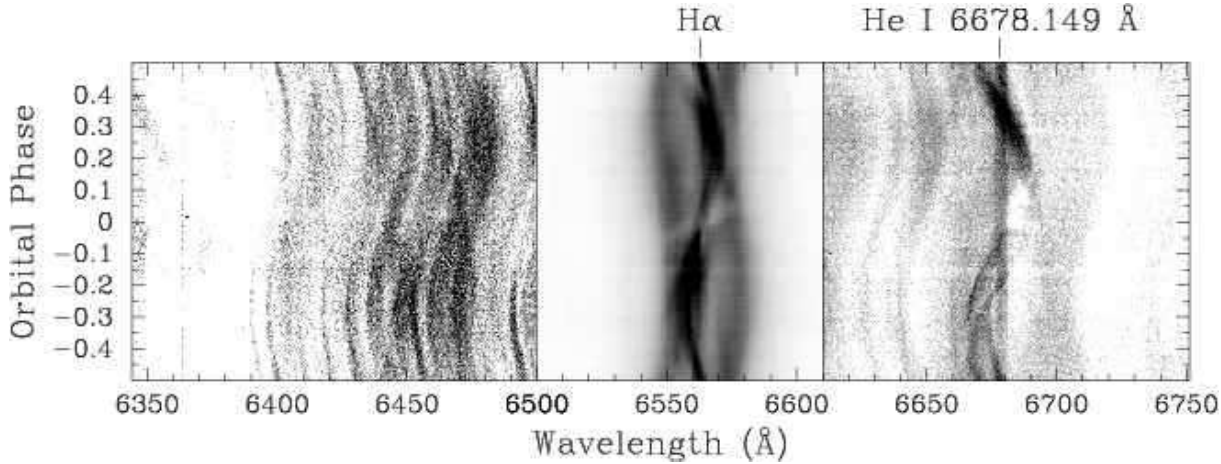}
\caption{\small The red spectra binned into 200 bins of
orbital phase. We reserve a panel to show the many features
seen in the continumm near H$\alpha$. The maximum plotted levels are:
1, 28 and 2 mJy above the continuum level for the continuum region,
\ha\ and \hei\ respectively.
\label{fig:reddata2001}}
\end{center}
\end{figure*}

Figures \ref{fig:bluedata2001} and \ref{fig:reddata2001} show the
average trailed spectra. Emission lines of \heii, \hb\ and \feii, \ha\
and \hei\ are visible. The trailed spectra show a remarkably rich mix
of features. Some of these are well-known from earlier work on U~Gem (Stover 1981;
Marsh et al 1990) and related systems; others are seen here for the
first time.

The light curves discussed in Section \ref{sec:light_curves} reveal
some flaring, more conspicuous in \hb. We looked for matches of this
behaviour in the trailed spectra, finding good correspondence in the
case of Balmer lines. The features are also stronger at high
velocities. This indicates that the inner accretion disc is the likely
source. We found no clear indication of these flares in \heii, \hei or
\feii.

\subsubsection{Standard features in the trailed spectra}

The lines show several well-know features of CV spectra such
double-peaked emission from the disc, an emission 'S'-wave from the
gas stream/disc impact point and emission from the mass donor star.
These occur with varying strength. The double-peaked emission from the
discs is best seen in \hb, while the emission from the donor star is
most obvious in \ha\ and \feii\ (Fig.~\ref{fig:bluedata2001}). The
latter line in particular shows that the emission from the donor star
fades around the time of eclipse, indicating that it is concentrated
on the side facing the white dwarf. This suggests that the emission is
triggered by irradiation from the accretion regions.

The bright-spot S-wave is especially clear in the light of \heii\
(Fig. \ref{fig:bluedata2001}) which has no visible disc contribution
at all. The \heii\ trail reveals that the S-wave itself is split into
two, with one broad component, but in addition a very narrow component
which executes a somewhat different path from its companion.. We
discuss these in Section \ref{sec:bright_spot} after we present
Doppler maps of our data. From the Doppler analysis we conclude that
both these S-waves come from the bright-spot region.

The effect of the partial eclipse on the lines is relatively subtle in
most lines, but is clearly seen in \ha\ as a region of low flux
running diagonally between the double-peaks from the lower-left
towards the upper-right between phases $0.95$ and $0.05$. This is the
effect of eclipsing Doppler-shifted emission from a prograde rotating
disc (Greenstein \& Kraft 1959; Young \& Schneider 1980). We will use
the eclipse in section~\ref{sec:inclination_angle} to measure
the orbital inclination of U Gem.

We now look at new features revealed by the high signal-to-noise of
our data.

\subsubsection{Low-velocity absorption and emission}
\label{sec:narrow}
\begin{figure}
\begin{center}
\includegraphics[width=3.5in,angle=270]{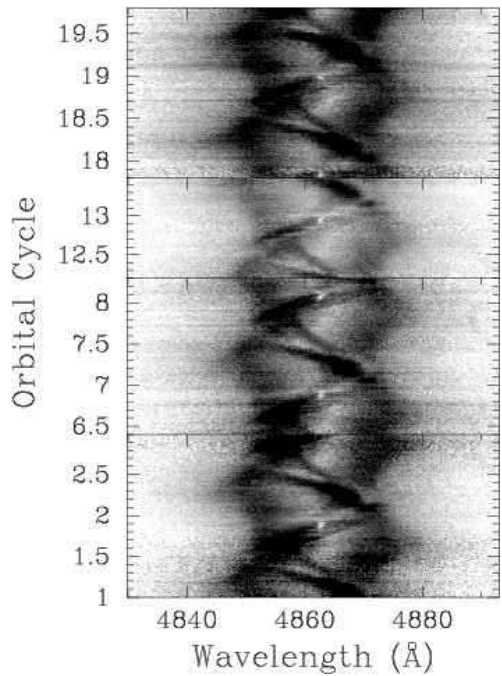}
\caption{\small The \hb\ data without phase-binning. Cycle 0 has been
set to the cycle 80740 according to the ephemeris used in this paper.}
\label{fig:hbeta2001}
\end{center}
\end{figure}

In \ha, but more obviously still in especially \hb\ (Figure
\ref{fig:hbeta2001}), a narrow absorption feature appears at line
centre immediately before eclipse between phases 0.91 and 0.98. To our
knowledge, nothing like this feature has been seen before in U~Gem or
any other system. It could be present now either because of the
improved data quality or because of a true change of structure in the
system.

Also in \ha, and more clearly in \hei, there is emission at low,
almost zero velocity, visible as a vertical line in \hei. We refer to
this as ``the low-velocity emission''.  Such emission is hard to
understand because there is no part of the system that is stationary
in this manner.



\subsubsection{Low-level structure in the continuum}

The continuum seen in Figure \ref{fig:reddata2001} shows a mass of
lines at a low level of flux. In several of these lines a modulation
is seen resembling the one produced by the mass donor's orbital
motion, so we attempted to track the source of the lines. To improve
the S/N of this region we phase binned the spectra in 50 phases. Then
we selected one of the spectra as a template and did a
cross-correlation with the set of phase binned data, backprojecting
later the result in what is called a skew-map (Smith, Dhillon \& Marsh
1998). This procedure showed that the features do indeed track the 
motion of the irradiated face of the donor, having the correct phase
and a semi-amplitude $K = 210\,\kms$ (see Section \ref{sec:shielding}).


We deduce that the features are line emission from the irradiated face 
of the donor. We could not make secure identifications for the lines,
although several match the wavelengths of FeII lines.

\subsection{Doppler tomography}
\label{sec:doppler}

\begin{figure*}
\begin{center}
\includegraphics[width=6.5in,angle=270]{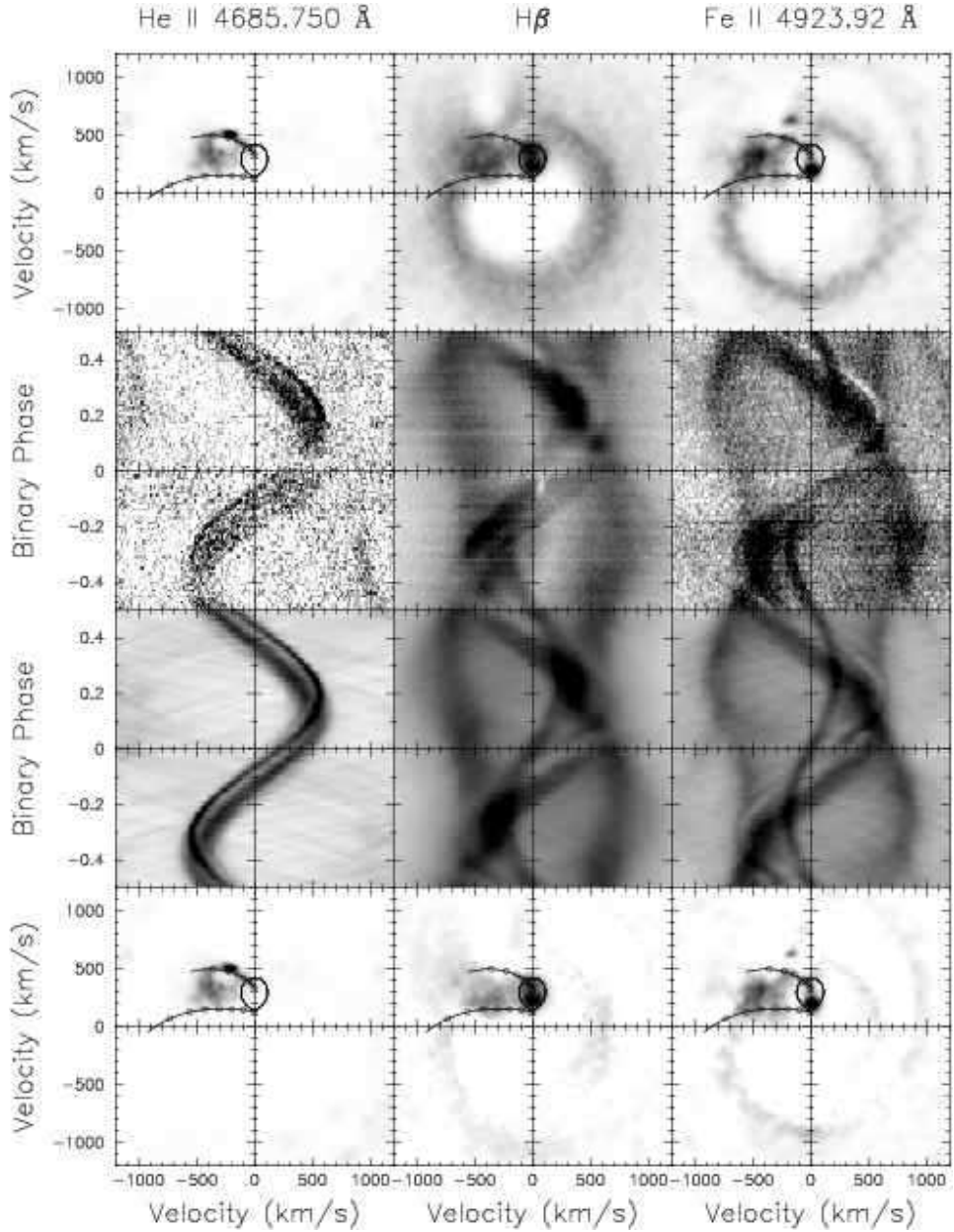}
\caption{\small Doppler tomography for the blue arm. From top to
bottom the rows show: 1) the Doppler tomograms, 2) the observed
spectra from which the tomograms were calculated, 3) synthetic data
computed from the tomograms, and 4) the tomograms after the
symmetrical part around $(0$,$-107.1)\kms$ was subtracted. The tomograms
include an outline of the Roche lobe of the mass donor and two
velocity trajectories for the stream. All scales start from zero with
uppermost levels chosen to improve the visualization of details. With
the exception of the third row (from top to bottom), which is evenly
saturated at 90 per cent of the peak flux, the first column is
saturated at 50 per cent, the second at 80 per cent, and the third at
60 per cent. See text for a thorough description of the features seen
in the tomograms.}
\label{fig:dopplermaps_blue}
\end{center}
\end{figure*}
To understand the features present in our trailed spectra, we used
Doppler tomography to study the emission lines. This technique
combines all the information observed in the trailed spectra and
generates a representation of the system in velocity coordinates. Due
to the current lack of understanding of how the system's velocity
coordinates should be translated to position coordinates, it is not
possible to get a direct picture of the binary star, and so the
interpretation is carried out in velocity coordinates. This allows a
visual separation of the contributions from different parts of the
binary star (e.g. the mass donor, the bright spot, the accretion disc,
etc.), and permits comparison with theoretical predictions under some
rough assumptions (e.g. Keplerian regime). For a thorough presentation
of the subject see Marsh \& Horne (1988).

\begin{figure*}
\begin{center}
\includegraphics[width=6.5in,angle=270]{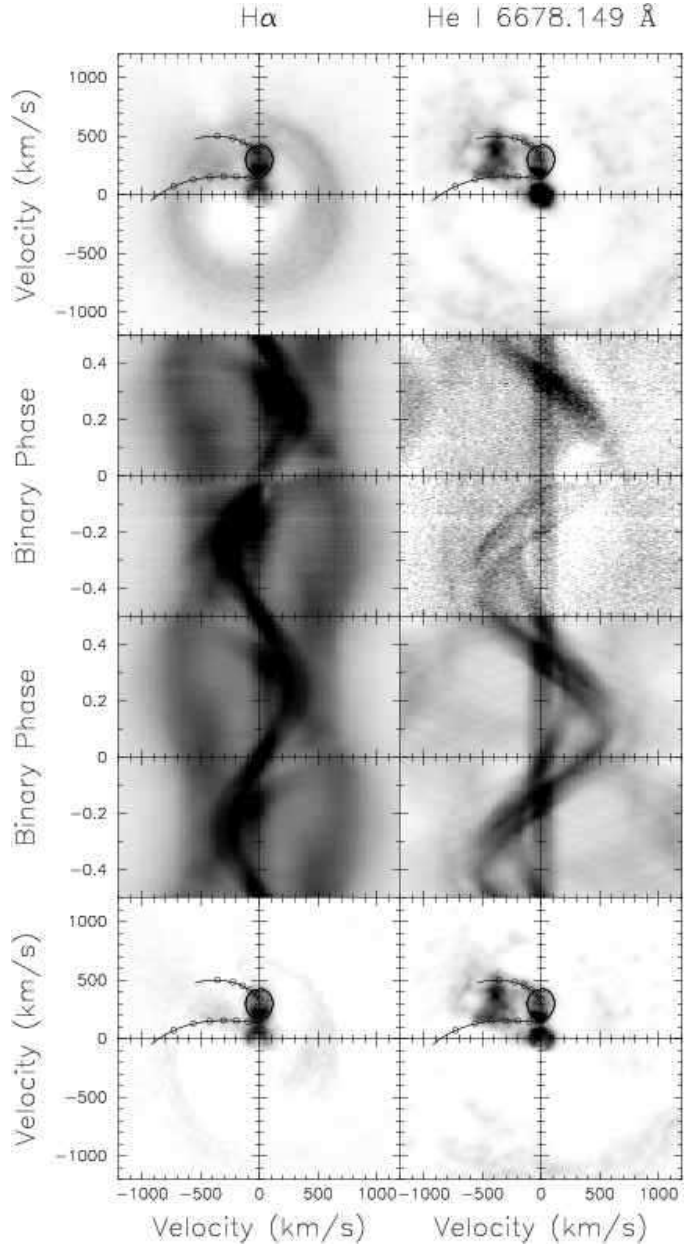}
\caption{\small Doppler tomography for the red arm. See
Fig.~\protect\ref{fig:dopplermaps_blue} for further details of the
display.  With the exception of the third row (from top to bottom),
which is evenly saturated at 90 per cent of the peak flux, the left
column is saturated at 80 per cent and the right column at 70 per
cent.}
\label{fig:dopplermaps_red}
\end{center}
\end{figure*}

%
%
In Figs.~\ref{fig:dopplermaps_blue} and \ref{fig:dopplermaps_red} we
present the tomograms for the data. The spectra taken during eclipse
were excluded from the computation of the maps, but they were included
as a template for the calculation of the synthetic spectra computed
back from the tomograms (third row from the top in the figures). We
plot the outline of the Roche lobe of the mass donor over the
tomograms. This was calculated by assuming $K_1=107.1\kms$ 
(Long \& Gilliland 1999) and $K_{2}=298\kms$. We
have already explained why $K_1$ is a good assumption. $K_{2}$ is a
combination of the three published values (Wade 1981, Friend et
al. 1990, Naylor et al. 2005), which are consistent with each other. 

%

We plot two velocity trajectories. The lower of the two is the
velocity of a ballistic gas stream. The upper one is the velocity of
the disc, assuming a Keplerian field, at positions along the
stream. The distance to the white dwarf is marked in these paths by
small circles at intervals of 0.1 times the distance of the inner
Lagrangian point ($R_{L1}$). Each trajectory is plotted from
1.0$R_{L1}$ to 0.4$R_{L1}$ (leftmost value). 

%

As expected from the discussion of the trailed spectra, neither the disc
nor the mass donor is detected in the \heii\ tomogram but the bright
spot is clearly resolved. In \hei\ the disc is almost undetectable,
but both the mass donor and the bright spot have strong
emission. Additionally, the \hei\ and \ha\ tomograms show a blob of
emission around the centre of mass of the system, the tomographic
equivalent of the ``low velocity emission'' we discussed earlier.  Both the disc and the
mass donor are clearly seen in \ha, \hb\ and \feii, the bright spot
appearing with varying degrees of strength.  The presence of the mass
donor in these tomograms is used later for a study of disc shielding
(Section \ref{sec:shielding}).

The tomogram for \feii\ displays two extra features. First, rather
like \heii, there is a very sharp spot seen above the path of
Keplerian disc velocities. Immediately below this spot there is a low
intensity region.  This corresponds to the absorption line 'S'-wave
seen in the trailed spectrum of this line
(Fig.~\ref{fig:dopplermaps_blue}). This is probably the absorption
equivalent of the sharp emission seen in \heii. However, because this
is only seen clearly over a very restricted range of phases ($0.15$ to
$0.45$), the emission spot gives a phase modulation -- in other words
it is likely to be an artefact and its displacement from the \heii\
emission spot is probably not significant. We conclude that both the
\heii\ emission and \feii\ absorption come from the same well-defined
structure associated with the gas stream impact. The spiral seen in
the upper-right quadrant of the \feii\ map is caused by the
contamination from a nearby line which can be seen on the right-hand
side of the trailed spectrum.

The bottom row in Figures \ref{fig:dopplermaps_blue} and
\ref{fig:dopplermaps_red} show the tomograms after we subtracted the
symmetrical part around $(0$,$-107.1)\kms$. Even after discarding the
contamination in \feii, we still see spiral structure in this map, and
also in \ha\ and \hb. This is further discussed in Section
\ref{sec:spirals}.

\subsubsection{Location of the bright spot}
\label{sec:bright_spot}

\begin{figure}
\begin{center}
\includegraphics[width=3in,angle=270]{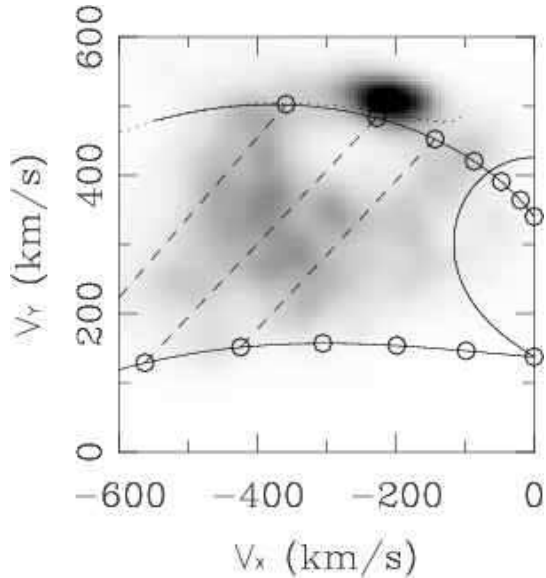}
\caption{\small Blow up of the \heii\ tomogram in the bright spot
region. The dashed lines connect the velocity of the ballistic gas
stream (black lower trajectory) and the velocity on the Keplerian disc
along the gas stream (black upper trajectory) for the same points at
distances of 0.4, 0.5 and 0.6 $R_{L1}$ to the white dwarf (right to
left). For the dotted trajectory see text. }
\label{fig:bright_spot}
\end{center}
\end{figure}

In addition to the sharp bright spot close to the Keplerian stream
trajectory that we see in the \heii\ map, we find a diffuse blob of
emission. This is located along the line that connects the sharp
feature to the matching spatial point that has the ballistic velocity
predicted for the stream (see Figure \ref{fig:bright_spot}), at
$\sim$0.5$R_{L1}$. The presence of the blob indicates that a mixture
of velocities coexist in the bright spot, the stream adjusting to the
velocities on the outer zones of the Keplerian disc (already noted by
Marsh et al.  1990). The value $\sim$0.5$R_{L1}$ (0.3 in units of the
orbital separation $a$) is consistent with observations by Smak (1984)
and is smaller than the radius of the tidally limited accretion disc
($0.44a$, see Warner 1995). 

Assuming that the position of the bright spot did not change
significantly during our observations, we can use the sharp feature
to estimate the bright spot's azimuthal location by treating the gas
stream/disc impact point as a point corotating like a rigid body with
the system. We conclude that the bright spot is face-on when the
system reaches phase $\sim$0.95, which is compatible with previous
results obtained from reconstructed light curves (e.g. Smak 1971).

The agreement between the position of the sharp feature and the
predicted path is not perfect, so we turn our attention to possible
causes. The path for the Keplerian disc velocities changes with
$K_{2}$, so one possibility is that a better value of $K_{2}$ is
needed, the problem being that measurements of $K_2$ are affected by
irradiation and the corrections needed are not entirely certain
(Friend et al 1990; Naylor et al 2005). We calculate that $K_{2}\sim
315\kms$ produces a good fit. This value for $K_{2}$ is not much
higher than the published values, i.e. $283\pm15\kms$ (Wade 1981),
$298\pm9\kms$ (Friend et al. 1990), $302\,\kms$ (Naylor et al
2005). Still, it is worth considering other possible causes.

The predicted stream path is based upon circular, Keplerian orbits
within the disc. However, the disc should be distorted by 3-body
effects, especially in its outer regions (Paczynski 1977). We computed
the velocity of the Keplerian disc along the path of the stream
allowing for such effects. The results are plotted as the dotted line
in Fig.~\ref{fig:bright_spot}, extending as far as the largest
non-intersecting orbit at $\sim$0.6$R_{L1}$. At radii smaller than
$\sim$0.4$R_{L1}$ the three body effects are negligible, but they do
alter the predicted velocities in the right direction, although not by
enough on their own.  Now however, $K_2$ need only be $\sim 308 \kms$
to fit, which is compatible with Wade's, Friend et al's and Naylor et al's
measurements.

Perhaps the most remarkable characteristic of the spot is how narrow
it is, being unresolved even at a resolution of $R \approx
10$,$000$. This, and the close match to the Keplerian disc velocity
along the stream, suggests that it must come from the disc immediately
prior to the stream/disk impact region. It could only do so by
irradiation from the main impact site.  This does not give a
corresponding feature at the ballistic stream velocity, possibly
because of the lower density of the stream. 

In contrast with the narrow spot in \heii, the other maps show a more
diffuse feature which lies between the two predicted stream paths.
Marsh et al (1990) ascribed this emission to the post-impact flow,
with gas taking a velocity intermediate between the ballistic stream
and the Keplerian disc. The extraordinary feature of this emission,
which was not clear at the lower resolution of Marsh et al's study, is
that it appears to extend all the way to the secondary star
(Figs~\ref{fig:dopplermaps_blue} and
Fig.~\protect\ref{fig:dopplermaps_red}). This should not be possible
because it would require the disc to extend well beyond its tidal
radius. There is a similar extension to high velocities which we will
discuss in following sections. We have no explanation for either of
these features.

\subsubsection{Low velocity emission}
\label{sec:low_velocity_emission}

In the case of the \ha\ Doppler map, the low velocity emission is
oddly asymmetric, displaying an elongated crescent-moon
(Fig.~\ref{fig:dopplermaps_red}). After subtracting the symmetrical
part around $(0$,$-107.1)\kms$ from the tomogram, the blob reveals
again a crescent shape. North et al. (2001) found a similar shape in
V426~Oph's low velocity emission.

Similar low velocity emission has been reported for other CVs.  A
summary of such reports is given in
Table~\ref{tab:low_velocity_emission}. There is no reason for any
component of the system to be at rest near the centre of mass, unless
the mass ratio $q$ has an extreme value. This is the case for example
for GP~Com for which $q \sim 0.02$ allows one to make the case for the
white dwarf to be responsible for the low-velocity emission it
displays (Morales-Rueda et al. 2003). In all the other cases the
emission is seen either at the centre of mass or somewhat displaced
towards the mass donor. Steeghs et al. (1996) suggested that the
emission was from gas originating in the mass donor, but trapped in a
prominence produced by a combination of magnetic, gravitational and
centrifugal forces within the rotating binary system. The prominence
would then be irradiated by the compact object and the disc, allowing
its detection as a component corotating as a rigid body with the mass
donor.

These ``slingshot prominences'', if similar to the prominences
observed in single stars, could be located up to several radii from
the magnetic star (e.g. Collier Cameron et al. 1990). If the
prominence behaves like the ones we observe on the Sun and rapidly
rotating stars like AB~Dor (see Collier Cameron et al. 1999), we
should expect them to be short-lived. Four nights should perhaps have
hinted at a variation either in flux or in size, but the feature is
consistent from night to night.  On the other hand, there is evidence
that loops on single stars could last $\sim$1 d ($\gamma$ Cas, Smith
et al. 1998). Perhaps under certain conditions prominences could last
for longer. 
%
There might be a common origin for the low velocity emission and the low
velocity absorption (section~\ref{sec:narrow}) if the gas trapped in the
prominences passes across the white dwarf at the time of eclipse, which seems
quite possible.

\begin{table*}
\caption{Summary of known Doppler tomographic reports of low velocity
emission. Under 'State', Q means 'quiescence', O means 'outburst', and
NA means 'not applicable'. WD stand for 'white dwarf' and S1996 for
'Steegs et al. (1996)'.}
\begin{tabular}{lllll}
System  &State&Lines &Suggested source &Reference\\
        &  &         &                                &\\
IP Peg  &O &\ha\     &Irradiated slingshot prominences&S1996\\
SS Cyg  &O &Balmer, \hei, \heii\  &Irradiated slingshot prominences&S1996  \\
AM Her  &NA&SI IV 1394\AA, NV 1239\AA&As in S1996 &G\"ansicke et al. (1998)\\
GP Com  &NA&\hei, \heii\ &WD (Morales-Rueda et al. 2003)  &Marsh (1999)\\
V426 Oph&Q &\ha\         &As in S1996 &North et al.(2001)    \\
U Gem   &Q &\ha, \hei\   &As in S1996 &This paper            \\
\end{tabular}
\label{tab:low_velocity_emission}
\end{table*}






\subsubsection{Stream-disc overflow?}
\label{sec:sd_overflow}

%
%
%

\begin{figure}
\begin{center}
\includegraphics[width=4in,angle=270]{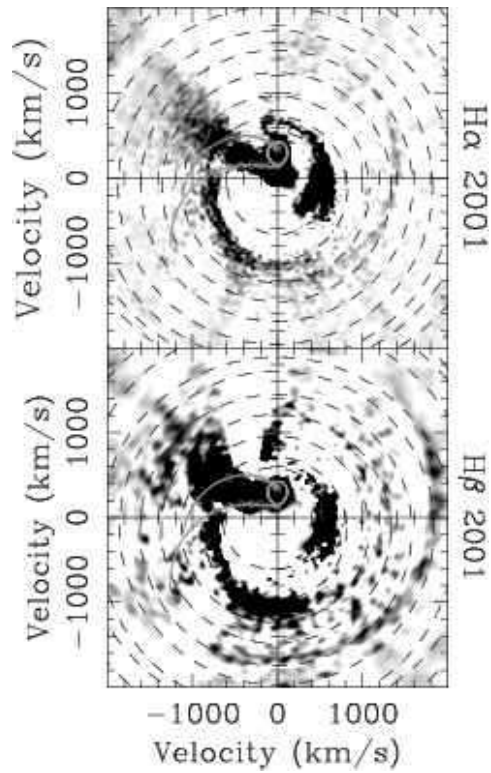}
\caption{\small \ha\ and \hb\ tomograms after subtraction of an
elliptical background. The ellipses used to interpolate the background
are plotted with dashed lines. The grey solid lines are the Keplerian
disc and ballistic stream velocities for the stream path. The tomograms
are saturated at only 3\%\ of the maximum level of the unsubtracted maps.}
\label{fig:continuum_subtracted}
\end{center}
\end{figure}

Simulations of the stream in U~Gem by Kunze et al. (2001) predict that
more than half the stream overflows the outer edge of the disc.  With
high signal-to-noise ratio and smooth disc profiles, our data are very
well suited to detection of such overflow which can be expected to
cause emisison along the path of the stream up to the point of closest
approach to the white dwarf (Livio et al. 1986; Armitage \& Livio
1998; Lubow 1989). In U~Gem the point of closest approach is
$\sim0.13\, R_{\rm L1}$ from the white dwarf, and in velocity space
corresponds to a region in the lower right quadrant of the Doppler
maps of Figs~\ref{fig:dopplermaps_blue} and \ref{fig:dopplermaps_red}.
To search more rigourously, we subtract the smooth background from the
disc by using an elliptical isophote fitting procedure described 
in Appendix~\ref{app:ellipses}. 
%
%
%
After performing the subtraction we
obtained the tomograms seen in Fig.~\ref{fig:continuum_subtracted},
which are scaled to only 3\% of the peak height of the raw
tomograms. The bright spot is clear in the maps, but there is no
evidence of stream-disc overflow along the path of the stream even at
this low level. In order to quantify this result we measured averages
and peak values over several samples of the region along which the
hypothetical overflow would travel, before and after the subtraction
of the elliptical continuum.  By comparing those values we conclude
that any possible overflow would contribute less than $3 \pm 2\%$ to
the flux from the disc surface.


Fig.~\ref{fig:continuum_subtracted} does show an excess extending up
and to the left of the main disc/stream impact region to well over
$1000\kms$. A similar effect was seen in EX~Dra (Billington \& Marsh 1996)
who suggested that non-kinematic broadening, such as Stark broadening,
could be responsible. 


\subsubsection{Spiral shocks in quiescence}
\label{sec:spirals}

\begin{figure*}
\begin{center}
\includegraphics[width=2.5in,angle=270]{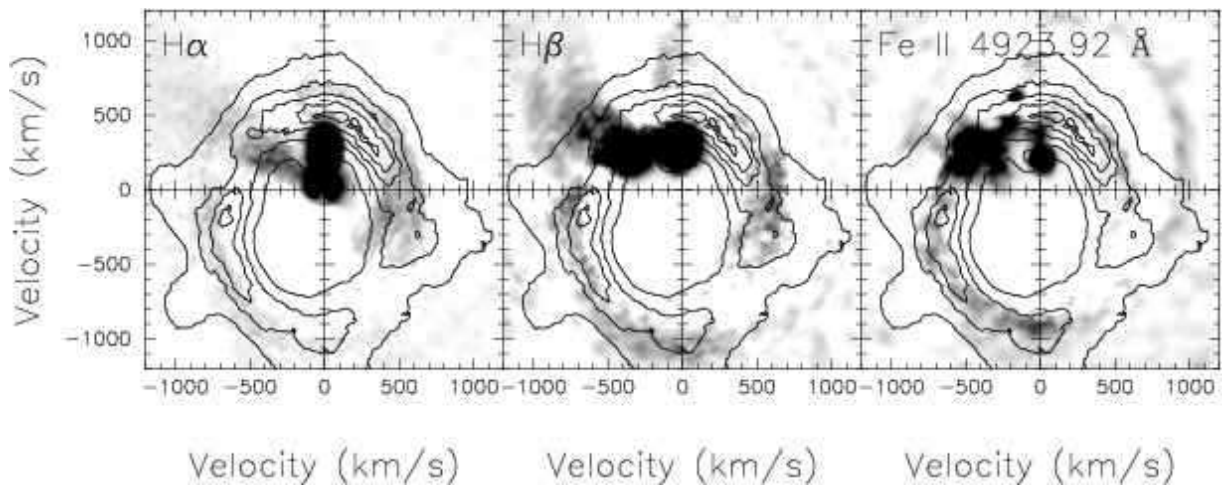}
\caption{\small Doppler tomograms for the Balmer lines and \feii\
after the symmetrical part around (0,-107.1)$\kms$ is subtracted.  The
tomograms are saturated at 20 per cent of peak level. Superimposed on
these tomograms we display isophotes (at 10, 30, 50, 70 and 90 per
cent of peak level) from a tomogram based on outburst observations of
U Gem (Steeghs, Morales-Rueda \& Groot, in preparation).}
\label{fig:spirals_quiescence_vs_outburst}
\end{center}
\end{figure*}

Figure \ref{fig:continuum_subtracted} reveals the presence of weak
spiral structure near the level of the continuum (we refer here to the
structure within $1000\,\kms$ of the centre of the images). In 
Figure~\ref{fig:spirals_quiescence_vs_outburst} we show
the maps with contours of spiral shocks observed during outburst 
(Steeghs, Morales-Rueda \& Groot in preparation) which
line up well with the asymmetries that we see in quiescence.
This spiral structure has been previously reported for U~Gem in outburst (Groot
2001), making it one of a handful of systems in which the phenomenon
has been detected. We emphasize that the amplitude of the spiral
structure we find here is \emph{far} less than that suggested by Neustroev and
Borisov (1998) for the same system. Their data shows pronounced
asymmetries, but were taken with poor spectral and phase resolution,
and most importantly, without complete orbital coverage. In contrast
our raw maps are very symmetric. We believe their claim of
strong spiral structure to be incorrect, and a result of the
difficulty of interpreting the complex variations of U~Gem with
inadequate orbital coverage.

The presence of spiral structure in the disc has been regarded as evidence of
``spiral shocks'' (Steeghs, Harlaftis \& Horne 1998), which had been predicted
in simulations of accretion discs (see Matsuda et al. 2000 for a review).  The
spiral shock interpretation requires the presence of large or unusually hot
discs. This prompted Smak (2001) and Ogilvie (2002) to propose alternatives to
the spiral shocks to explain the phenomenon. They suggest that the spiral
features revealed by Doppler tomography could be explained as the consequence of
3-body effects (Paczynski 1977). Ogilvie extends the model of Paczynski from 2
to 3 dimensions, allowing for a complete set of physical conditions in his
model. He concludes that, due to tidal distortions, some regions of the disc
would thicken, later being irradiated by the white dwarf.  The pattern of
thickening would be slightly spiral, but no waves or shocks would be involved in
the process. In Doppler tomography the pattern would show up as the kind of open
spiral structure seen in published tomograms.

These two competing explanations predict a different evolution of the
spiral structures from outburst to quiescence. The spiral shock model
requires hot material to have open spirals. In quiescence, the spirals, 
if present, are expected to be relatively tightly wound (Steeghs and Stehle 1999).
%
Assuming that the asymmetries that we see are indeed related to the outburst
spiral structure, our observations show little change in morphology from
outburst to quiescence, i.e. the spirals do not appear tightly wound. This
argues against the spiral shock model as an explanation of the spiral structure
detected in U~Gem and favours Ogilvie's (2002) and Smak's (2001) explanation.
%
%

Although similar patterns are seen in several lines, we remain cautious about
claiming a definitive detection of spiral structure in the quiescent U~Gem. We
emphasize again that the maps are really extremely symmetrical, and given the
presence of strong bright-spot components which have strengths modulated in
orbital phase, it is hard to rule out the possibility that we are seeing
low-level artefacts.  It is also worth noting that the size of the disc we find
from the bright-spot of $\sim$0.3a (Section \ref{sec:bright_spot}) is close to
the minimum observed radius of U~Gem's disc (Smak 1984) and well inside the
tidal radius where one would expect Ogilvie's and Smak's model to operate, let
alone tidally-driven spiral shocks.


\subsubsection{Disc shielding}
\label{sec:shielding}

\begin{figure*}
\begin{center}
\includegraphics[width=2.5in,angle=270]{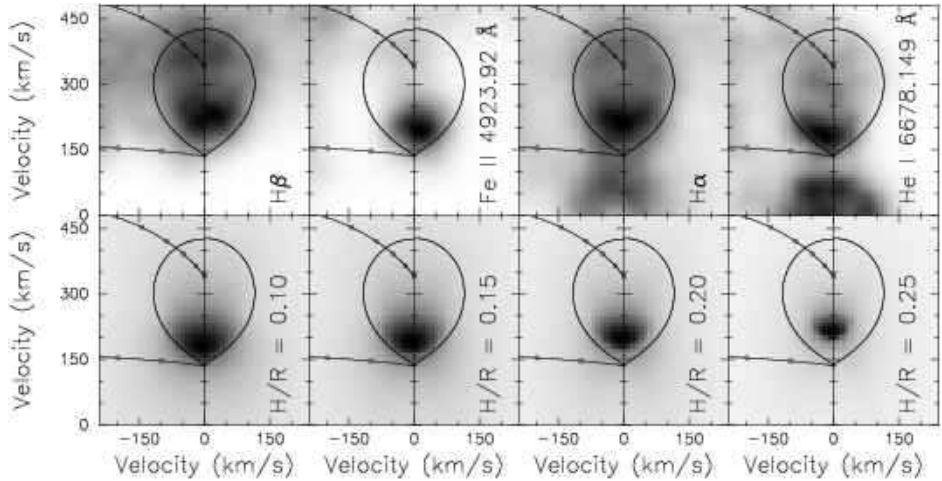}
\caption{\small Real and simulated disc shielding. The top row is a
close view of those tomograms centered on the mass donor. The lower
portion of the Roche lobe is not fully irradiated, which is explained
by shielding of the mass donor's equatorial regions due to an
effective height of the disc's outer rim. The bottom row shows
simulations for different H/R ratios. All the images are saturated at
100 per cent, the colour scales starting at zero flux.}
\label{fig:disc_shielding}
\end{center}
\end{figure*}

A notable feature in most of the tomograms, in particular in \feii, is
the mass donor's irradiated face. Small sections of these maps,
centered on the donor star, are presented in the top row of Figure
\ref{fig:disc_shielding}. The figure shows that the
irradiation-induced emission avoids the equator of the mass donor,
indicative of shielding by the disc (Harlaftis 1999; Morales-Rueda et
al. 2000). The irradiation in \ha\ and \hb\ seems to be located
further from the L1 point than for the other lines. This is suggestive
of a changing effective thickness due to wavelength dependent opacity
at the EUV wavelengths needed for photoionisation of the elements
involved. This has been previously seen in other CVs (Morales-Rueda et
al. 2000).

In the bottom row of Figure~\ref{fig:disc_shielding} we show simulated
tomograms for a system with the same parameters of U Gem but with
varying height-to-radius ratio (H/R) in the disc. We included some
broadening in the simulations by convolving the synthetic datasets
with a gaussian with FWHM $15\,\kms$. Values of H/R from 0.10 to 0.25
were considered (increasing by 0.05 each time).  For $H/R > 0.30$, the
height completely prevents the irradiation to take place.  We estimate
from this that the H/R ratio for U Gem is $\sim$0.15 (\hei),
$\sim$0.20 (\feii) and $\sim$0.25 (\ha\ and \hb). These values could
be a little smaller if $K_2$ is larger than we have assumed (raising
the Roche lobes of Figure~\ref{fig:disc_shielding}), as suggested by
the sharp spot in HeII (section~\ref{sec:bright_spot}).

\subsection{Inclination angle}
\label{sec:inclination_angle}

\begin{figure}
\begin{center}
\includegraphics[width=4in,angle=270]{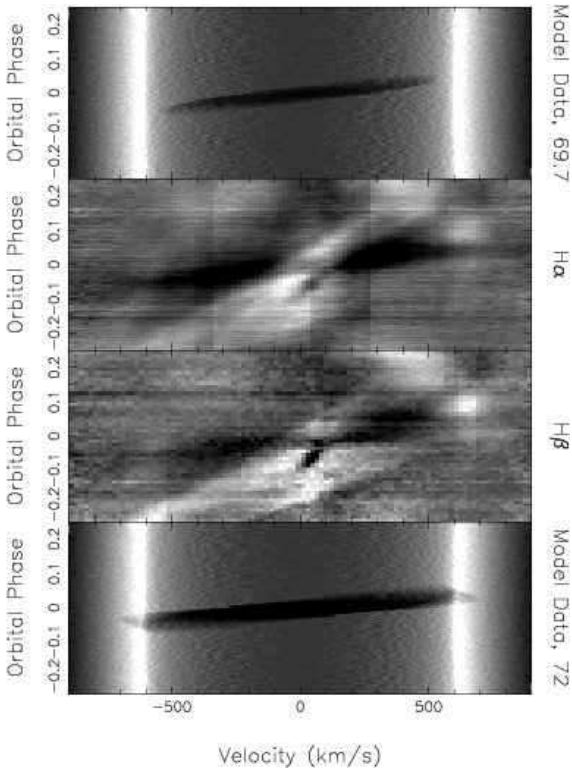}
\caption{\small The top and bottom panel show simulated data of the eclipse of
the emission lines for
$69.7^{\circ}$ and $72^{\circ}$ respectively. The central panels show
data from the Balmer lines. The central panels were
processed to improve visibility of the extension of the eclipse (see
text for details). All the panels are saturated at 100 per cent. Note that the
data are affected by emission components not included in our simulations.}
\label{fig:eclipses}
\end{center}
\end{figure}

Zhang \& Robinson (1987) derived an orbital inclination angle $i = 69.7^\circ$
for U~Gem from a detailed lightcurve fitting model. Other estimates of $i$ have
displayed a trend to somewhat smaller values, but mainly in the region of
$65^{\circ} < i < 70^{\circ}$.
%
%
Given the radial velocity semi-amplitiudes, the inclination determines the
component masses. Long (2000) and more recently Naylor et al (2005) encountered
a problem when combining a measurement of the gravitational redshift of the
white dwarf, which fixes $M/R$, with other constraints including inclination
estimates.  The problem, summarised
in Fig.~8 of Naylor et al. is that the gravitational redshift gives a rather
small radius for a given mass, which only crosses the Hamada-Salpeter
mass-radius relationship for $M_1 ~ 1.25 \, \mathrm{M}_\odot$ implying at the
same time a low inclination of $i \sim 65^\circ$ and a radius well below the
value predicted from U~Gem's UV spectrum and parallax (Long 2000; Harrison 1999).
%
The Hamada-Salpeter relation and the radius measured from the UV flux are in
agreement for a much higher inclination of $72$ to $74^\circ$ and a lower white
dwarf mass, $\sim 1.05\,\mathrm{M}_\odot$ (Naylor et al. 2005), but then one must
suppose that the gravitational redshift, a combination of careful measurements
from the UV (Long \& Gilliland 1999) and I-band (Naylor et al. 2005) to be
seriously in error.

In this section we use the eclipse of the emission lines, and specifically the
highest velocity in the lines that is eclipsed to provide a new independent
constraint on the orbital inclination. We find a high inclination favouring the
low-mass solution for the white dwarf. The key point is that in a partially
eclipsing system, since the eclipse does not reach the white dwarf, there will
be a velocity in the lines above which there is no eclipse. If we can determine
this point, we can measure the inclination. We do so by simulating the eclipse
and comparing the simulation with the data (Figure \ref{fig:eclipses}). The top
panel shows a simulated trailed spectra when $i$ is set to $69.7^{\circ}$, with
$K_1$ and $K_{2}$ set to the values mentioned before. The three central panels
show real data after removal of the mean spectrum. The extension of the
eclipse in these panels easily reaches $700\,\kms$ while in the simulated data
of the top panel it does not go beyond $550\,\kms$.  The inclination must be
larger than $69.7^\circ$; we favour a value of $i \sim 72^\circ$ (lowest
panel, Figure \ref{fig:eclipses}).

%
The emission line eclipse strongly favours the low-mass, high inclination
solution and is in conflict with the measurements of gravitational redshift.
Our data allows us to check the systemic-velocity of the donor star which
is one component of the gravitational redshift. We did this by subtracting from
the dataset the orbital motion corresponding to the centroid of the mass donor's
irradiated region. Then we averaged the spectra in phase ranges for which the separation
between the signal from the irradiated region and other contaminating signals
(bright spot, disc) was maximum. This produced well defined peaks to which
we fitted gaussian solutions. We calculated the systemic-velocity of the donor
as the offset of the centre of the gaussians with respect to their wavelengths
at rest. The result is  $36\pm2\kms$ (average of \ha, \hb\ and \feii). This is
compatible with $29\pm6\kms$ from Naylor et al. (2005) but, being higher,
it reduces a little the value of the redshift. However, this reduction is far
from enough to resolve the discrepancy between the two mass solutions
for U~Gem.

\subsection{Orbital velocity of the white dwarf} \label{sec:rv}
\begin{figure}
\begin{center}
\includegraphics[width=3in,angle=270]{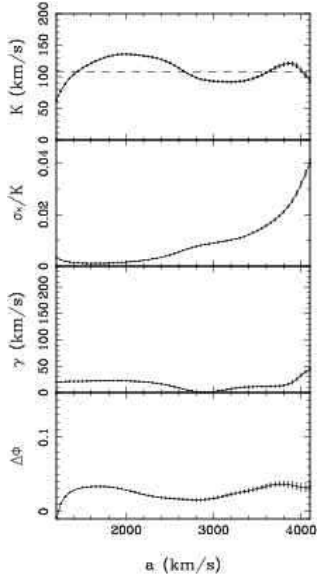}
\caption{\small The diagnostic diagram for \ha. A separation of
gaussians equal to $\sim$3400 km $s^{-1}$ seems appropriate to avoid
most of the jump in noise close to 4000 km $s^{-1}$ in the
$\sigma_{K}$/$K$ plot. This separation corresponds to a plateau in the
$K$ plot, from which the value $\sim$95 km $s^{-1}$ is read for
$K_1$.  The dashed horizontal line in the top panel marks the level
107.1 km $s^{-1}$.}
\label{fig:rv_diag_halpha}
\end{center}
\end{figure}
\begin{figure}
\begin{center}
\includegraphics[width=3in,angle=270]{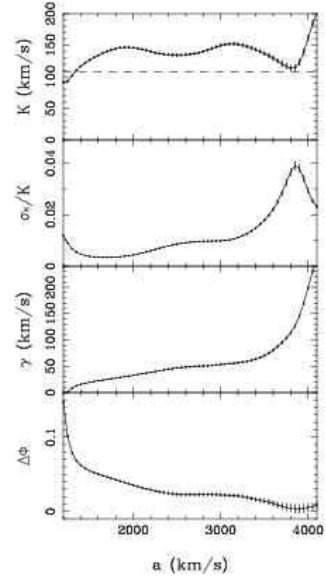}
\caption{\small The diagnostic diagram for \hb. A separation of 3000
or 3200 km $s^{-1}$ seems appropriate in this case. This points to a
value of 150 km $s^{-1}$ for $K_1$.  The dashed horizontal line in
the top panel marks the level 107.1 km $s^{-1}$.}
\label{fig:rv_diag_hbeta}
\end{center}
\end{figure}
\begin{figure}
\begin{center}
\includegraphics[width=5in,angle=270]{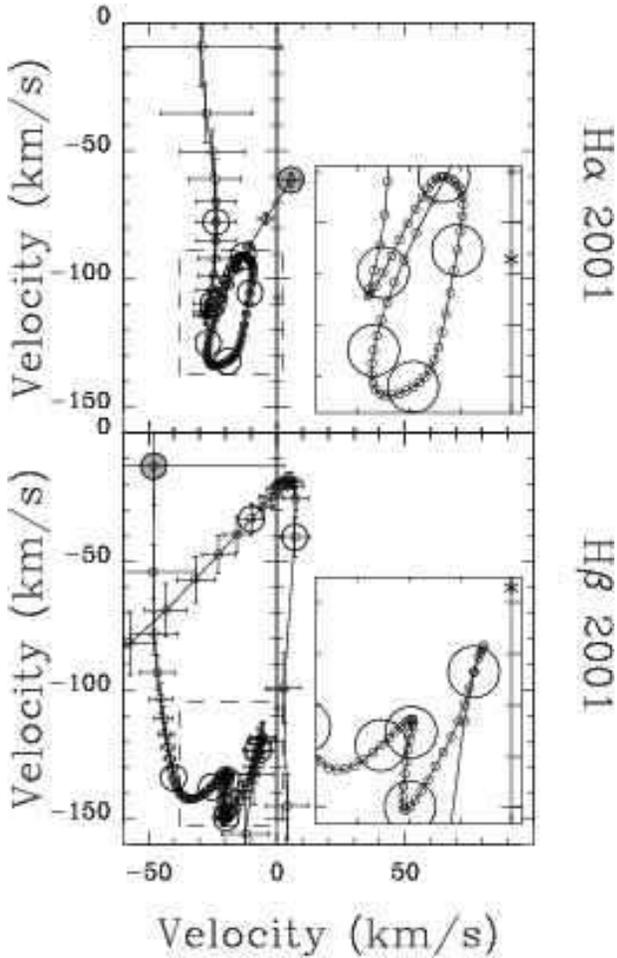}
\caption{\small The result of applying our new method to \ha\ and
\hb. See text for details of how to read the plots. A certain trend of
the points in direction to (0,107.1) km $s^{-1}$ is seen twice in the
\hb\ plot.  \ha\ shows no clear trend towards (0,107.1) km $s^{-1}$.
}
\label{fig:rv_new_method}
\end{center}
\end{figure}

The value of $K_1$ for U~Gem is known accurately thanks to a direct
observation of metal lines from the white dwarf's surface (Long \&
Gilliland, 1999). The result, $107.1 \pm 2.1 \,\kms$, can be used to
test the effectiveness of the techniques already in use, which are
mostly based on assuming that the emission lines originated on the
accretion disc will provide information from which the motion of the
white dwarf can be tracked.

Two traditional techniques to measure $K_1$ are the diagnostic
diagram (Shafter, Szkody \& Thorstensen 1986) and the light-centre
diagram (Marsh 1988). Both use the double-Gaussian method (Schneider
\& Young 1980; hereafter SY80) to measure the radial velocities from a
dataset. This method consists in convolving the continuum subtracted
line with two Gaussians of identical width and varying separations.

%
%
In the diagnostic diagram technique, the velocities measured are then
fitted with an orbital solution $V = \gamma - K \sin(2 \pi [\phi -
\Delta\phi])$, and the parameters of the fit are plotted in a diagram
like those presented in Figs.  \ref{fig:rv_diag_halpha} and
\ref{fig:rv_diag_hbeta}. $K_1$ is considered to coincide with $K$ when
$K$ is stable over a range of Gaussian separations (a in the plot) or
at the point just before $\sigma_K/K$ increases sharply. In
Figs. \ref{fig:rv_diag_halpha} and \ref{fig:rv_diag_hbeta} we see that
this happens for a = $3400\,\kms$ giving values of $K_1$ = 95 and
$150\,\kms$ for \ha\ and \hb\ respectively.  The plots in the
diagnostic diagrams display curves that behave only partially as
expected. In particular there is no simultaneous convergence
towards final values in the panels (compare, for instance, top and
bottom panel of Figure \ref{fig:rv_diag_halpha}).

In the light-centre technique, the radial velocities measured using
the double-Gaussian method are plotted in velocity space and the line
of points obtained is extrapolated towards the $K_y$ axis. The value
for $K_1$ is the point at which the extrapolated line crosses $K_y$.
The plot obtained, not reproduced here, is very similar to
Figure~\ref{fig:rv_new_method}, which is described below.

\subsubsection{A new method to measure $K_1$}

%
%
A problem with the double gaussian measurement is that each spectrum is treated
one by one, before the individual radial velocities are fitted by a sinusoid.
The process breaks down as soon as the signal-to-noise in one of the spectra
becomes too low for the measurements to be made. We can do somewhat better by
fitting all spectra simultaneously, still employing the double-gaussian method
because of its isolation of particular velocities in the line profiles. Denoting
the velocity, flux and uncertainty of the $i$-th pixel of the $j$-th spectrum by 
$v_{ji}$, $f_{ji}$ and $\sigma_{ji}$ respectively, we therefore miminise the 
following function:
\begin{equation} \label{eq:chi2}
\sum_{j=1}^{N} w (\mathcal{V}_{ji})
                              \left[
                              \sum_{i=1}^{M} f_{ji} g(\mathcal{V}_{ji})
                              \right]^{2}
\end{equation}
where $\mathcal{V}_{ji}$ is the velocity of a pixel relative to
the velocity of the spectrum $U_j$,
\begin{equation} \label{eq:defined_variable}
\mathcal{V}_{ji}= v_{ji}-U_{j},
\end{equation}
$w$ is a weighting function given by
\begin{equation} \label{eq:w}
w(\mathcal{V}_{ji}) =
\frac{1}{\sum_{j=1}^{M}\sigma_{ji}^{2}g^{2}(\mathcal{V}_{ji})},
\end{equation}
and $g$ represents two gaussians of standard deviation $\sigma_G$ separated
by $a_G$,
\begin{equation} \label{eq:g}
g(v) =  \exp\left(-\frac{\left(v -
\frac{1}{2}a_{G}\right)^2}{2\sigma_{G}^2}\right)
   - \exp\left(-\frac{\left(v +
\frac{1}{2}a_{G}\right)^2}{2\sigma_{G}^2}\right)
\end{equation}
The function of Eq.~\ref{eq:chi2} is minimised by fitting the velocities of the
spectra $U_j$, $j = 1$ to $N$. If these quantities were fitted independently, then we are back
to the standard double gaussian method for one spectrum which finds the velocity
$U$ for which $\sum_i f_i g(v_i - U) = 0$. The difference is that we do not allow the $U_j$
to be independent, but instead constrain them to represent a circular orbit
as so:
\begin{equation} \label{eq:U}
U_{j} =
\gamma - K_{x}\cos(2\pi\phi_{j}) + K_{y}\sin(2\pi\phi_{j}),
\end{equation}
fitting instead the standard orbital parameters $\gamma$, $K_x$ and $K_y$.
Uncertainties were estimated by a bootstrap procedure (Diaconis \& Efron 1983),
with 1000 bootstrap samples being chosen from the original set of spectra.

%

The end products are $K_x$ and $K_y$ values as a function of gaussian
separation, exactly as with the old procedure. The gain is greater robustness in
the presence of noise.


Figure \ref{fig:rv_new_method} shows the application of this new method for
performing the analysis of radial velocities. We use a small cross to mark the
point with coordinates (0,-107.1) $\kms$, the direct measurement for $K_1$ from
Long \& Gilliland (1999). The points in the diagrams show the variation with
separation for two gaussians of FWHM $=100\,\kms$. The initial separation is
$1200\,\kms$ and is marked with a large grey circle. Small circles mark steps of
$50\,\kms$ while large circles marks steps of $500\,\kms$. In both panels an
inset with a blow up of the region close to (0,-107.1) $\kms$ has been included.
In the insets we have ommited the error bars.

Although the \hb\ plot does on two occasions shows trends towards the known
value,  it would be impossible in the absence of Long \& Gillilands's value
to deduce it from our measurements, although they are at least in its vicinity.
This should not come as a suprise; the diagnostic diagrams have a phase offset
from zero, and whatever causes the distortion will distort the new method as well.
The chief conclusion is that even far in the wings, there is something that
distorts the radial amplitudes by of the order $20$ to $50\,\kms$, and that
we cannot improve on Stover's (1981) estimate of $137 \pm 8 \,\kms$.

We carried out several simulations to see what could disturb the wing velocities
using the model applied in Section~\ref{sec:inclination_angle}. We began the
simulations considering just an accretion disc and an opaque secondary star. In
this way we obtained double-peaked phase-shifted spectra in which an eclipse by
the secondary star was included but no flux was contributed by it to the
spectra. We found that both techniques were able to accurately recover the
correct value for $K_1$ from this dataset.  We then added a nearby line so
that the separation and relative fluxes of both lines would approximate the
situation of having \hei\ or \feii\ close to \ha\ or \hb\ respectively. Again
both methods correctly extracted $K_1$ from the data. Then we added a bright
spot (its parameters tuned to fit the feature observed in our data) to the main
line (\ha-like and \hb-like) and found that the measurements were seriously
distorted by its presence. However, after increasing the gaussian separation
beyond $\sim2000 \kms$ the methods again reliably extracted $K_1$ from the
data. Lastly, we added a bright spot to the companion lines (\hei-like and
\feii-like) and found that the recovery of $K_1$ was no longer possible. 

In summary, we cannot recover U Gem's $K_1$ to better than $\sim 40 \,\kms$,
and we suspect that the problem is caused by the need to go far into
the line wings to avoid the bright-spot, which brings with it the problem
of contamination by the bright-spot emission of nearby lines.

\section{Conclusions}
\label{sec:discussion}

We have presented high spectral resolution and signal-to-noise optical
spectra of U~Gem with the aim of testing the use of the broad emission
lines in tracing the motion of the accreting white dwarf in CVs. U~Gem
is particularly well suited for this task because the true motion of
its white dwarf has been measured from {\it HST} spectra by Long et al
(1999).

Despite observations on a $4$m-class telescope of one the brightest dwarf novae,
we find values of the semi-amplitude of the white dwarf ranging from 80 to $\sim
150 \,\kms$ which contain not a hint that the correct value is the $107 \,\kms$
measured by Long et al (1999). Neither the diagnostic diagram (Shafter et al
1986) nor the light centre technique (Marsh 1988) were of any help. The root
cause of the problem seems to be the bright-spot which causes an asymmetry that
extends to large velocities in Doppler maps, and which is present in the same
way in pairs of close spectral lines (\ha\ and \hei; \hb\ and \feii), thus
violating the usual assumption of a trend towards symmetry at high velocity.

Our data reveal a number of new phenomena for which we can offer only tentative
explanations. These include low velocity absorption and emission that we suggest
imply the presence of prominences on the mass donor star.

U~Gem shows a complex structure in the region of the gas stream/impact
region. Extended emission lies between the velocity expected for the
ballistic gas stream directly and the velocity of the Keplerian disc
along the gas stream. This emission extends both to very high and very
low velocities in the maps. Most surprising of all is the presence of
a narrow component of the bright-spot in \heii, which appears to be
unresolved even at $R \approx 10$,$000$. This narrow spot appears to
come from the disc rather than the stream, but seems to require a
slightly larger value of radial velocity amplitude for the secondary
star than has been measured to date.

We find no evidence of stream overflow, but we find weak spiral asymmetry in the
(quiescent) disc, similar in nature to those seen during outburst. This may
provide support for Ogilvie's and Smak's explanation for the presence
of spiral structure based upon 3-body effects.

We find an inclination angle of $\sim72^{\circ}$. This supports a solution
for the mass of U~Gem's white dwarf in agreement with the Hamada-Salpeter
mass-radius relationship and with the radius measured from the UV flux.
This reinforces the discrepancy first noted by Naylor et al. (2005) regarding
the mass estimated from gravitational redshift. This constitutes an 
important open puzzle.

\section*{Acknowledgments}

EU was supported by PPARC (UK) and Fundaci\'on Andes (Chile) under
the program Gemini PPARC-Andes throughout most of this research.
LMR was supported under a PPARC PDRA grant and TRM under a PPARC SRF 
during some of the period over which this work was undertaken. 

Danny Steeghs' comments about sections of this article greatly 
improved the final version. We also wish to thank him for his Doppler 
map of U~Gem in outburst.

The authors acknowledge the data analysis facilities provided by the
Starlink Project which is run by the University of Southampton on
behalf of PPARC.

This research has made use of NASA's Astrophysics Data System 
Bibliographic Services, as well as of the wonderful resources provided
by Hartley Library at the University of Southampton.

During the long time that took the writing of this article, Soledad 
Mart\'inez Labr\'in had to use many times her incredible ability to
make starry the darkest nights. EU is lovingly indebted to her.

\appendix

\section{Ellipse fitting of Doppler map isophotes} \label{app:ellipses}

In a Doppler tomogram is usually possible to find at least two closed
contours at a particular flux level. This happens because the tomogram
is usually dominated by the disc continuum, in such a way that the
flux increases from the centre of the map towards a ring of maximum
intensity, the flux then decreasing from this ring outwards. This
scenario makes the traditional methods to fit ellipses from the field
of galactic surface photometry unstable (e.g. Kent 1983, Jedrzejewski
1987), because in them most levels occur only once. On the other hand,
the techniques employed in computer vision, for the characterization
of conics in an image, are very suitable for application here. We
devised a new strategy by dividing the problem in two stages:
\begin{enumerate}
\item Selection of points pertaining to a closed contour, at a given
level of flux.
\item Fitting of an ellipse to such points.
\end{enumerate}
\subsection{First stage}
Tracing the contours at a certain level and extracting the points pertaining to
a single closed contour at the same level are two completely different problems.
The former is readily solved in a purely graphical manner (e.g.  Burke 1987),
while the latter (ours) is non-trivial and, as far as we know, not certainly
solvable without human assistance. Consequently the first stage admits several
approaches varying in efficiency. We devised one in which from many angular
directions either zero or one pixel was chosen, according to its proximity to
the pixel selected in the previous direction and to the general pattern of the
contours (e.g. no crossed contours were allowed).

The procedure could be carried out almost on its own in the case of a
smooth continuum, but it required close supervision when dealing with
finely detailed (or noisy) structure. The main danger is for the
algorithm not to close the contour or to take a wrong detour before
coming back to the expected path.
\subsection{Second stage}
Once all the points for a particular level were chosen, we fitted them
with ellipses according to an analytical ellipse-specific method
proposed by Fitzgibbon, Pilu \& Fischer (1999). The task is
accomplished via a very efficient matrix procedure. In a non-matrix
form we can explain the procedure as the fitting of a general conic of
the form:
\begin{equation} \label{eq:general_conic}
Ax^2+Bxy+Cy^2+Dx+Ey+F=0
\end{equation}
but adding the constraint:
\begin{equation} \label{eq:constraint}
4AC-B^2=1
\end{equation}
so that the result will specifically be an ellipse. The constraint
adds an arbitrary factor of scaling to the solution, but the algorithm
provides for the removal of the effect at a later stage. The resulting
general coefficients $A$, $B$, $C$, $D$, $E$ and $F$ were converted to the
more meaningful values $V_{xc}$,$V_{yc}$ (coordinates of the centre of
the ellipse), $a$, $b$ (size of the two semiaxes), $e$ (eccentricity)
and $\theta$ (angle of rotation). Uncertainties were estimated by a
bootstrap procedure over the points.

The solution given by the method of Fitzgibbon et al. (1999) has been
proved to be unique, but it must be noted that it has a certain bias
to low eccentricities, specially if only a few points are fed to the
algorithm. This is fine for our purposes because we use many points
for the fitting (basically limited by the resolution of the map) and
also we do not want unrealistically high values for $e$ (which for an
accretion disc we expect to be close to 0 for most isophotes). If this
was a problem, the bias could be removed by another bootstrap
procedure at the cost of computing time (see, for instance, Cabrera \&
Meer 1996).

\begin{figure*}
\begin{center}
\includegraphics[width=2.5in,angle=270]{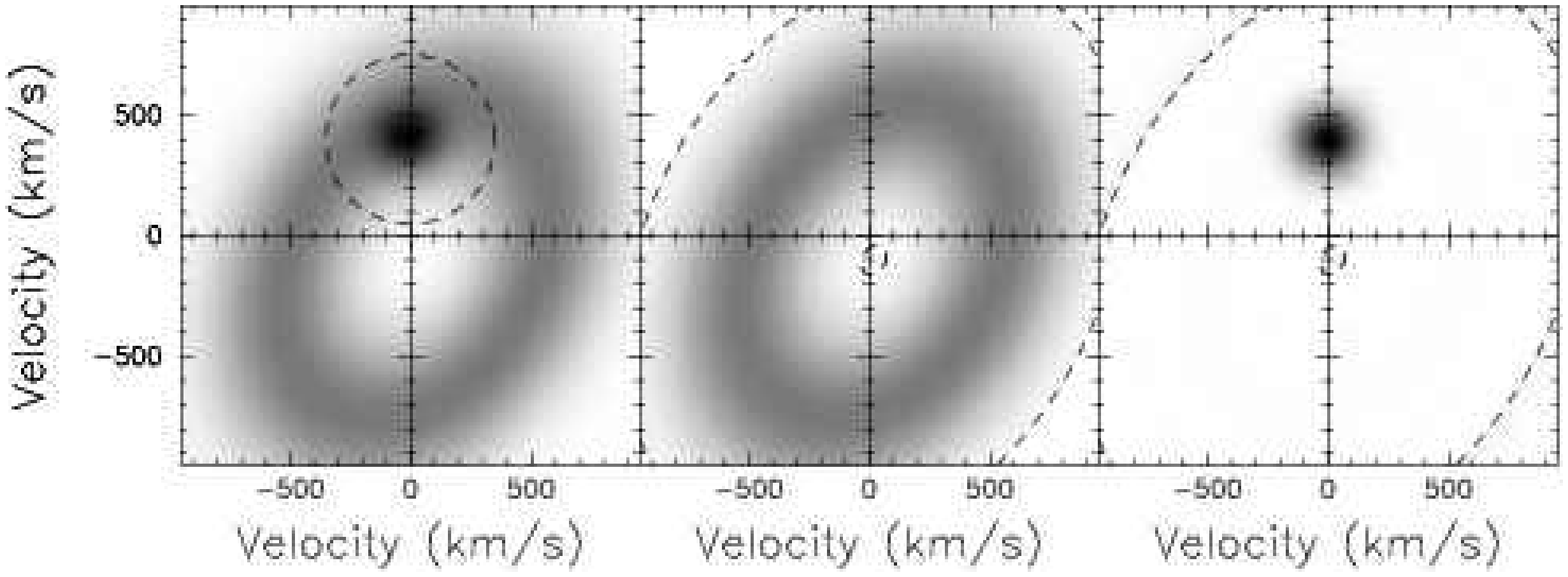}
\caption{\small Performance of the elliptical background subtraction. The left
panel shows a synthetic tomogram with an elliptical background and a
spot of enhanced emission. The region inside the dashed circle was
masked during the fitting of ellipses. The middle panel shows the
continuum interpolated from the fitted ellipses. In both the middle
and right panel the dashed ellipses are the innermost and outermost
fitted ones. The right panel shows the result of subtracting the
middle panel from the left one. The (correct) 2-dimensional Gaussian
nature of the synthetic spot is made more evident in the right panel
than it was in the original figure, in which it seemed to be elongated
along the curvature of the disc.}
\label{fig:test_subtraction}
\end{center}
\end{figure*}

\subsection{Application: subtraction of elliptical background}

By subtracting the disk from the original Doppler map we can greatly improve the
visibility of asymmetrical features. Although we expect disc isophotes to have
low values of $e$, this value is generally not be exactly 0. We therefore used
the results from the ellipse-fitting process described above to subtract an
interpolated elliptical background. To do this, we found for every point the two
closest ellipses and then we linearly interpolated their fluxes according to the
distance of the point to both ellipses. Special cases were: a point exactly in
the path of an ellipse (its flux was used), and points beyond the region fitted
(no flux was assigned).

In Figure (\ref{fig:test_subtraction}) we show an example of the
performance of the method by using a synthetic Doppler map.

\label{lastpage}

\end{document}